# Synthesis of $Ti_2B_2Cl_x$ MBenes in molten salts from theoretical and experimental perspectives

Rodrigo M. Ronchi [1], Emile Defoy [2], Andrejs Petruhins [1], Justinas Palisaitis [3], Lianghao Yu [1], Lan Tang [4], Solenn Reguer [5], Dominique Thiaudière [5], Ningjun Chen [1], Durga Sankar Vavilapalli [1], David Portehault [2], Jonas Björk [1], Per O. Å. Persson [3], Johanna Rosen [1]

[1] Materials Design division, Department of Physics, Chemistry, and Biology (IFM), Linköping University, SE-581 83 Linköping, Sweden.

[2] Sorbonne Université, CNRS, Laboratoire de Chimie de la Matière Condensée de Paris (CMCP), 4 place Jussieu, F-75005, Paris, France

[3] Thin Film Physics division, Department of Physics, Chemistry, and Biology (IFM), Linköping University, SE-581 83 Linköping, Sweden.

[4] School of Mechanics and Optoelectronic Physics, Anhui University of Science and Technology, Huainan 232001, China

[5] Synchrotron SOLEIL, L'Orme des Merisiers Saint-Aubin BP 48, 91192 Gif-sur-Yvette, France

**Corresponding authors: johanna.rosen@liu.se*

**ABSTRACT**

The unique properties and application possibilities of two-dimensional (2D) materials motivates the exploration of different nanolaminated compounds. Here, by using a molten salt approach, we selectively etch $Ti_2InB_2$ with $ZnCl_2$ to produce a multilayer (ml) $Ti_2B_2Cl_x$ MBene. Scanning transmission electron microscopy, in combination with energy dispersive X-ray, and electron energy loss spectroscopies show that In atoms are completely removed from the precursor upon etching, being replaced by chlorine surface terminations with a coverage $1.1 < x < 1.4$. Further, *in situ* X-ray diffraction indicates a direct biphasic transformation from $Ti_2InB_2$ to ml-MBene, with no signs of intermediate phase formation. A computational framework based on density functional theory further corroborates these experimental observations by showing a negative reaction free energy for the formation of ml-MBene, favourable over all competing processes. In addition, A-element substitution into to the 3D $Ti_2ZnB_2$ phase is predicted to be endergonic, consistent with the absence of experimental evidence for its formation. Initial Li-ion battery performance evaluation showed a stable discharge capacity similar or better than MAX phases and other borides. Altogether, the theoretical framework combined with materials synthesis and characterization provides a general approach for 2D materials development, for further expansion of the family of 2D materials.

## INTRODUCTION

After the discovery of graphene [1], research on two-dimensional (2D) materials has expanded significantly because of their unique properties derived from their intrinsic high surface area-to-volume ratio. After graphene, other 2D materials were discovered, including hexagonal boron nitride [2], transition metal dichalcogenides [3,4] and, in 2011, MXenes [5]. These different 2D materials hold great potential for numerous applications, including energy storage, catalysis, gas separation, and water desalination and purification [6–8]. Among them, 2D transition metal borides have recently attracted increasing attention due to their predicted metallic conductivity, tunable surface chemistry, accessible active sites, and low diffusion barriers, which make them particularly promising for applications such as battery electrodes, electrocatalysis, and energy storage systems [9–11].

Efforts to discover and synthesize 2D borides have focused on the synthesis of metal–boron (M–B) stacks by selectively removing the A-element (e.g., Al or In) from precursor boride compounds known as MAB phases. Based on the similarities with so-called MAX phases (the parent materials for 2D MXenes), different experimental attempts to etch these materials have been performed. In 2021, hydrofluoric (HF) acid was used to selectively etch Al and Y from the $(Mo_{2/3}Y_{1/3})_2AlB_2$ *i*-MAB phase, resulting in a 2D $Mo_{4/3}B_{2-x}$ MBene [9] that showed potential for hydrogen evolution reaction (HER) [11]. More recently, 2D $TiBT_x$ (T= Cl) MBenes were tested in lithium-ion batteries [10].

Since the first realization of MBenes [9], various selective etching approaches have been explored, including hydrochloric [12–15] and hydrofluoric acids [9,16] and, more recently, Lewis acidic molten salts (MS) [16–19]. However, most strategies have struggled to produce both multilayers and single sheets of 2D MBenes. For example, etching MoAlB only partially removed Al, yielding $Mo_2AlB_2$, another MAB phase, regardless of whether MS [17] or acids [12,14] were used. MS-etching of $Hf_2InB_2$ led to complete oxidation (i.e., HfBO), unlike halogen-terminated MXenes obtained through similar methods [16]. Additionally, acid etching failed to remove the A-layer for $Fe_2AlB_2$ [9], $Mo_5SiB_2$ [9], $Ti_2InB_2$ [20] and $Hf_2InB_2$ [16]. Furthermore, although a dealloying reaction could remove In from $Ti_2InB_2$, this was accompanied by a structural reorganization resulting in a 3D TiB crystal structure, instead of a multilayer stack of 2D sheets [20]. These experimental studies demonstrated that the synthesis of 2D MBenes is significantly more challenging compared to MXenes [21].

Here, we report the selective etching of $Ti_2InB_2$ into a ml-$Ti_2B_2Cl_x$ MBene. By using the Lewis acidic molten salt $ZnCl_2$, indium atoms could be completely removed from the MAB phase precursor, accompanied by functionalization of the ml-boride sheet surfaces with chlorine terminations. *In situ* XRD experiments indicated a direct biphasic transformation from $Ti_2InB_2$ to ml-MBene, somewhat different from MXene reactions, in which the reaction sometimes occurs through an intermediate phase [22]. In addition, a rational framework for this pathway could be obtained by employing a recently developed theoretical methodology [23], where ml-$Ti_2B_2Cl_2$ exhibited the lowest (negative) reaction free energy among possible pathways, while the intermediate compound ($Ti_2ZnB_2$) had positive reaction free energy. This integrated experimental–computational strategy provides general insights into selective etching mechanisms and offers a predictive framework for designing and understanding the formation of 2D materials derived from MAB phases and other layered compounds.

## RESULTS AND DISCUSSION

$Ti_2InB_2$ was prepared by a solid-state reaction of elemental powders with excess amounts of both Ti and In elements. X-ray diffraction (XRD) analysis of the as-synthesized sample (Fig. S1) shows the presence of the targeted phase ($Ti_2InB_2$) together with impurities, including $TiB_2$, $Ti_{2.24}In_{1.76}$, $Ti_3In$ and $Ti_3In_4$, as previously reported [20]. Upon cleaning with diluted hydrochloric acid (2 mol $L^{-1}$), the Ti–In phases were dissolved, leaving only $Ti_2InB_2$ and $TiB_2$. The phase content, determined by Rietveld refinement, was 90 wt% $Ti_2InB_2$ and 10 wt% $TiB_2$. These numbers are close to those reported in the original study (93.7 and 6.3 wt% [20], respectively), despite the drastically reduced annealing time used for synthesis (6h vs 6 days [20]). The corresponding refinement and structural parameters are shown in Fig. S2 and Table S1, respectively. SEM analysis of the sample after HCl cleaning revealed a close-packed layered morphology (Fig. S3) with the atomic molar ratio of In:Ti confirmed by EDX to be 0.46, consistent with the expected value of 0.5. Boron was not quantified due to the limited accuracy of EDX towards light elements, but its presence was qualitatively indicated in the EDX maps shown in Fig. S4.

Selective etching of the indium layer in $Ti_2InB_2$ was performed using the MS approach, schematically illustrated in Figure 1a. From the different synthesis conditions evaluated (Fig. S5), the optimized MS etching conditions were 5 hours at 600 °C using 4:1 salt-to-MAB molar ratio, significantly smaller than that recently used (10:1) [10]. Figure 1b compares the XRD patterns of the $Ti_2InB_2$ precursor after HCl cleaning and the etched material, termed 'ml-MBene'. The characteristic peaks of the MAB phase disappear after MS reaction, while a low-angle peak appears at 2θ=9.25°, corresponding to an enlarged (002) plane d-spacing value of ≈9.5 Å, compared to ≈7.9 Å of $Ti_2InB_2$.

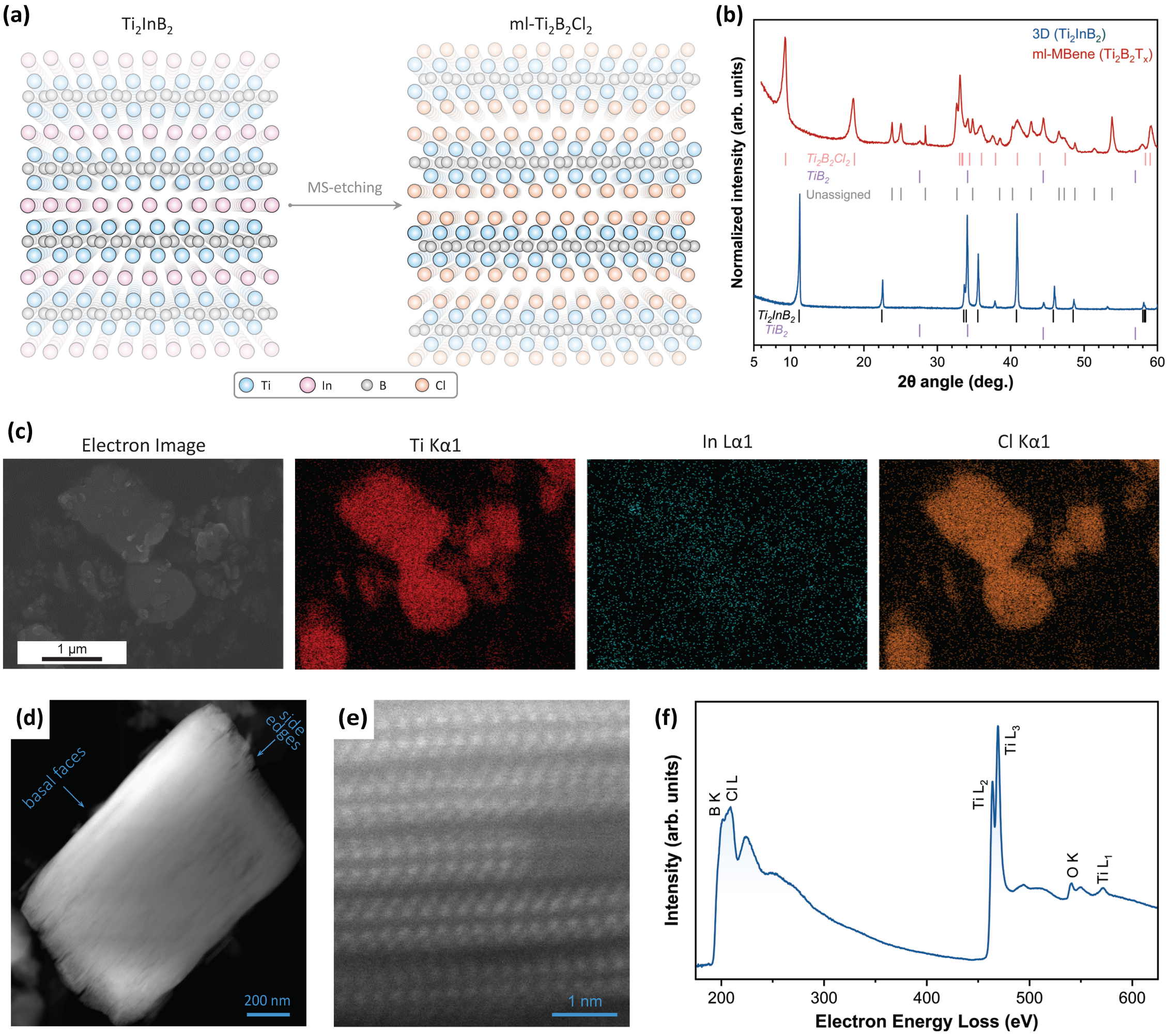

Figure 1: (a) Schematic conversion of $Ti_2InB_2$ into the ml-MBene $Ti_2B_2Cl_2$. (b-e) Characterization of the ml-MBene obtained after MS etching. (b) powder XRD, (c) SEM image with Ti, In and Cl EDX maps, (d) overview and (e) high-resolution HAADF-STEM images, and (f) EELS spectrum.

Besides the dominant reflections associated with the ml-MBene compound and $TiB_2$ impurity inherited from the parent phase, additional weak reflections can be observed after the molten-salt etching, including those at 2θ ≈ 23.8°, 25.1°, 28.4°, and 53.8°. These features are likely associated with minor secondary phases formed during the etching and/or washing process. Comparison with ICSD database entries suggests that some of these peaks may be related to titanium oxide phases, such as $TiBO_3$, $TiO_2$, and $Ti_2O_3$, although definitive assignment is hindered by peak overlap and the relatively low intensity of these reflections. Nevertheless, these minor features do not affect the primary conclusion regarding successful In extraction and formation of ml-$Ti_2B_2T_x$.

While SEM did not indicate significant changes in the particle morphology after reaction in the molten salt and washing (see Fig. S3), EDX elemental maps (Figure 1c and Fig. S4) and survey spectra statistics shown in Fig. S6 suggested complete removal of In accompanied by the incorporation of Cl into the particles. The atomic ratio after etching was 1 Ti: 0.00 In: 0.55 Cl: 0.06 Zn, compared to 1 Ti: 0.46 In in the 3D precursor. The multilayer nature of the synthesized ml-MBene can be seen in the overview HAADF-STEM (Figure 1d) and the atomically resolved high-resolution image shown in Figure 1e. This image exhibits atomic number-dependent contrast, i.e., Ti atoms have the brightest contrast, Cl atoms are less bright, and B ones are too light to be resolved. The Ti–B layered backbone is the same as in the $Ti_2InB_2$ precursor, but the boride layers are now separated by two layers of terminating atoms, resulting in an increased *c* lattice parameter (19.1 Å). STEM-EDX measurements corroborate SEM-EDX. An EELS spectrum acquired at the center of the particle basal face (i.e., top and bottom surfaces in Figure 1d) showed the presence of Ti, B and Cl (Figure 1f), as revealed by their characteristic edges (note that the B K- and Cl L-edges are close in energy and partially overlap). The Ti: B: Cl molar ratio retrieved from EELS data was 1: 0.9: 0.6, corresponding to the chemical formula $Ti_2B_{1.8}Cl_{1.2}$. Notably, boron was quantified with a ratio close to the expected nominal ratio (Ti:B = 1).

SEM-EDX indicated the presence of oxygen in ml-particles (Fig. S4 and Fig S6). To better understand its presence and evaluate the change of chemical composition at the surface, high-resolution STEM-EDX maps and line profiles were acquired. Figs. S7a-b clearly indicate the presence of oxygen at the particle side edges, with depths of approximately 20–30 nm. EELS quantification at the side edge showed a Ti: B: Cl: O molar ratio of 1: 0.9: 0.0: 1.0 ($Ti_2B_{1.8}O_2$), indicating that this region has only oxygen and no chlorine atoms. This observation contrasts with EELS analysis at the center of the basal face, with negligible amounts of oxygen. Thus, the ml-MBene particle shows an oxide-rich shell along the sheet edges.

From the results discussed above and summarized in Tables S2 and S3, we conclude that the ml-MBene composition is $Ti_2B_2Cl_x$ ($1.1<x<1.4$), with chlorine terminations mostly on the basal faces of the ml-particles, while they are replaced by oxygen on the side edges (Figs. S8). Both the non-ideal surface coverage of halogen atoms as well as the edges oxidation are in line with previous reports on MXenes [24,25] and recent $Ti_2B_2Cl_2$ synthesis [26], compiled in Table S4. In the case of halogen-terminated MXenes, incorporating oxygen atoms into the vacant termination sites has been theoretically predicted to be viable [27]. That theoretical work [27] provides insights into our experimental observation that both oxygen and chlorine terminations are identified, consistent with oxygen atoms on MS-etched MXene surfaces, suggested to originate from the washing process [24].

Finally, the MS-etching process showed a high yield, which is an important parameter for scalable purposes. The calculated normalized conversion yield from $Ti_2InB_2$ to $Ti_2B_2Cl_{1.5}$ was approximately 75%, demonstrating an efficient 3D to 2D conversion (see more details in SI). When full coverage is assumed ($Ti_2B_2Cl_2$), the calculated yield (68%) would be similar to the one previously reported for $Ti_3AlC_2$ conversion into $Ti_3C_2Cl_2$ MXene (67%) [24]. While the present study focuses on ml-MBenes, it should be noted that using tetramethylammonium hydroxide (TMAOH), the material could be delaminated (Fig. S9), in line with ref [26].

It is important to note that the environment in which the synthesis takes place clearly influences the MS reaction. Fig. S10 compares MS-etching experiments performed in sealed glass tubes and with open crucibles. While the sealed tube sample with 4:1 salt-to-MAB molar ratio shows complete removal of $Ti_2InB_2$ XRD peaks, the open crucible experiment in the same conditions shows still significant reflections of the MAB precursor. A similar trend is also observed by decreasing the salt-to-MAB ratio to 3:1, evidencing that the open crucibles provide less reactive experimental conditions. This may be attributed to the high volatility of $ZnCl_2$ at 600°C combined with gas flow in the furnace, which eliminates the salt from the reaction medium and, therefore, decreases its reactivity. Importantly, irrespective of the condition used, the formation of ml-MBene samples is always observed, only the extent of the etching is modified.

The results shown above confirm the successful synthesis of a ml-MBene with Cl terminations. For MAX phases, it has been suggested that the ability of a Lewis acid to withdraw electrons from an A-element is related to the electrochemical redox potentials, i.e. etching of an A element would be achieved by using a Lewis acid with a higher (less negative) redox potential [25]. For example, at 700 °C, the calculated redox potentials of $Al^{3+}/Al$ and $Cu^{2+}/Cu$ versus $Cl_2/Cl^-$ are -1.85 and -0.43 V, respectively. The higher redox potential of $Cu^{2+}/Cu$ would indicate that $CuCl_2$ salt would be an effective Lewis acid to etch Al-based MAX phases [25]. However, this approach fails to describe the MS etching of the boride achieved in the present work, as the calculated redox potential of $Zn^{2+}/Zn$ (-1.44 V) is lower than for the $In^{3+}/In$ couple (-1.08 V), which would suggest that etching is unfavorable. This discrepancy may be associated with the fact that redox potentials are calculated by considering A in its elemental state, which may differ considerably from the actual state of A atoms within a compound, such as a MAB phase. In addition, it only indicates that the salt would react with the ternary precursor, but does not provide information on the phase formed, i.e., whether MBenes, replacement of the A-layer, or even other competing phases like binary borides.

To provide a more accurate insight into the driving force for etching, the reaction free energies of the MS reactions were estimated using both first principles calculations and tabulated values [23]. Four scenarios are explicitly compared: (i) A-element substitution, where the A-element of the MAB phase is replaced by the metal element from the salt; (ii) ml-MBene formation, in which the A-element is etched while halogens from the salt bind to the ml-MBene surfaces, (iii) complete disintegration, i.e., the MAB phase fully decomposes into salts and/or its constituting elements; and (iv) 3D boride formation, where the A-element is etched, and the ml-MBene layers collapse into a 3D boride. The essence of the method is to construct process-specific phase diagrams for different reactions, allowing the comparison of competing processes for different salt concentrations and at different temperatures. A schematic representation of these four processes is shown in Figure 2a.

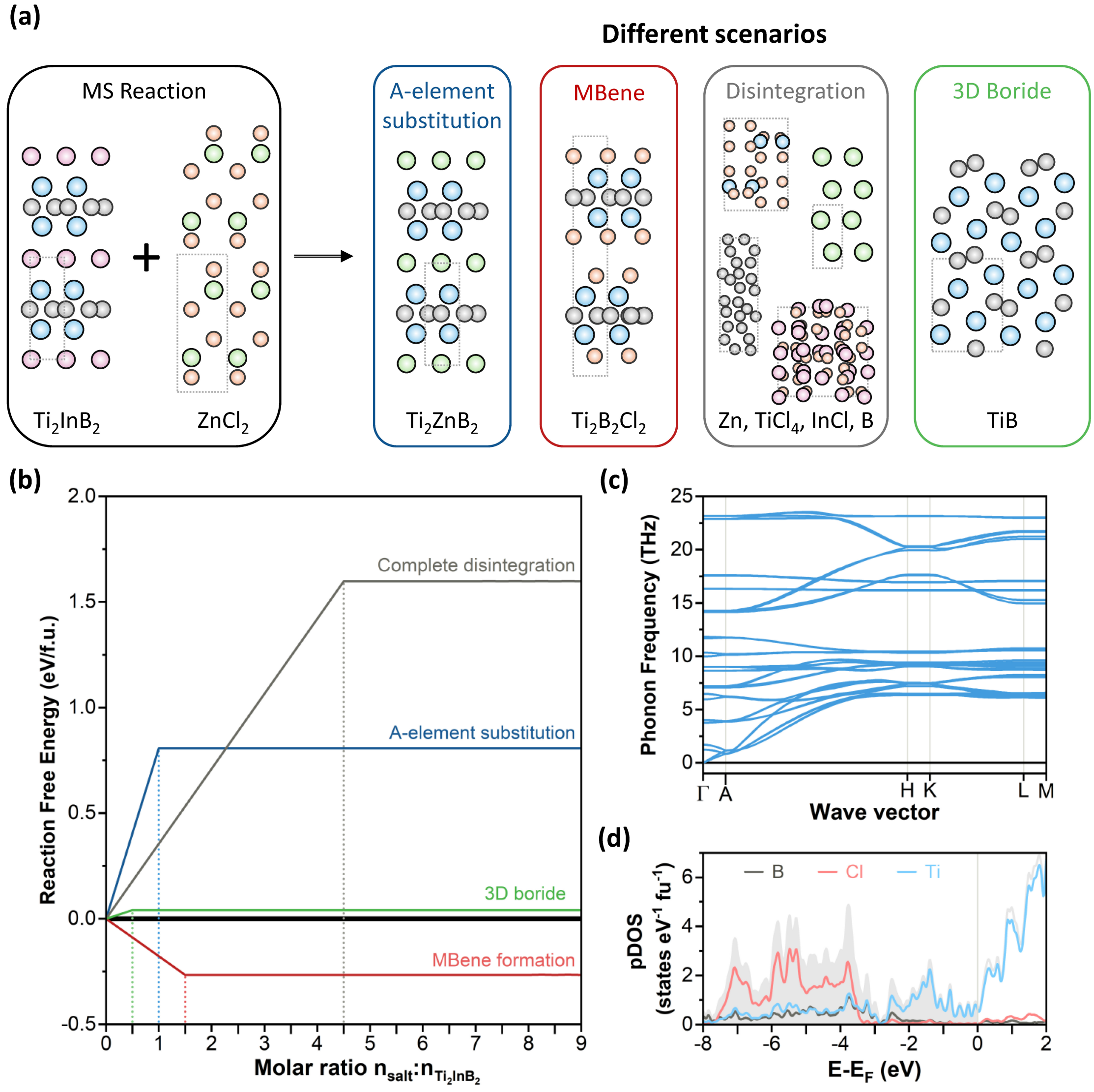

Figure 2: (a) Schematics of reaction pathways, showing the different possible scenarios considered. (b) Comparison of reaction Gibbs free energies as a function of the salt-to-MAB molar ratio. The vertical dotted lines represent the molar ratio at which each process is completed (before the threshold, only a fraction of the MAB has reacted). (c) Calculated ml-$Ti_2B_2Cl_2$ phonon spectrum and (d) projected density of states.

The multilayer ml-MBene structure used for DFT calculations was constructed based on the experimental XRD and STEM results. As a first approximation, we used stoichiometric composition ($Ti_2B_2Cl_2$), avoiding calculating vibrational energies of disordered systems (with vacancies and/or oxygen atoms). Based on previous indications for MBenes and MXenes [27–29], Cl atoms were positioned above/below one B site (i.e., hollow positions with respect to the metal surface), while each $Ti_2B_2Cl_2$ layer is misaligned with the layer above in an A-B-A stacking sequence (see 'CIF file' in SI). After relaxation, the calculated lattice parameters were a=b=3.13 and c=18.95 Å, consistent with STEM (3.05 and 19.1 Å) and XRD (3.12 and 19.06 Å) measurements. This good agreement between experimental and theoretical results verifies our findings and supports the use of the non-local vdW-rev-DF2 potential [30] to account for the weak van der Waals interactions between the layers [31,32].

Figure 2b presents the calculated process-specific phase diagram comparing the calculated reaction free energies for the four processes involved in etching of $Ti_2InB_2$ with $ZnCl_2$ at varying salt-to-MAB molar ratios, for a temperature of 900 K. The specific reactions and energies can be seen in SI. Notably, the MBene formation has the lowest reaction free energy across all salt concentrations considered and is the only exergonic pathway, making it the sole thermodynamically viable reaction among those considered. This result is in contrast with the comparison of redox potentials of the metals in their elemental form [25] and supports our methodology for screening MAB-to-MBene reactions in MS-etching.

The present theoretical analysis is formulated within a thermodynamic framework and is therefore designed to capture equilibrium driving forces rather than kinetic effects such as energy barriers. Thus, the predictions should be interpreted as general trends under idealized conditions, rather than as quantitative descriptors of specific experimental protocols. In this context, experimental variables, such as performing reactions in sealed tubes or open crucibles, can influence the salt concentration at which MBene formation occurs, but ml-MBene formation was observed in both cases. The strong agreement between thermodynamic predictions and experimental observations underscores the robustness of the framework and supports its extension, originally developed for MAX phases, to MAB phases and potentially other layered compounds.

The calculated phonon spectra for the 2D-ml ml-MBene structure are shown in Figure 2c. There are no negative (imaginary) frequencies, indicating the dynamic stability of the structure. The projected density of states (pDOS) for the ml-MBene structure suggests metallic behavior (see Figure 2d). This was further confirmed experimentally by the absence of absorption edges in UV-Vis spectroscopy measurements (Figure S11). While energies close to the Fermi level ($E_F$) are dominated by Ti d states, Cl and B states appear more significantly from energies lower than -3 eV vs. $E_F$. These pDOS characteristics are similar to what has been found previously for halogenated MXenes [27]. Important to note is that the degree of surface termination coverage is expected to strongly influence the density of states at the Fermi level (see ml-$Ti_2B_2$ pDOS in Figure S12) [33]. Lower termination coverage ($x < 2$) for MXenes increased the number of non-bonding metal states at the Fermi level due to a decrease in the number of M–T bonds [27]. Thus, a similar effect may be expected for ml-MBenes, increasing the states at the Fermi level and, as a result, potentially increasing the electrical conduction. In addition, Figure S12 shows that the density of states of fully terminated ml-$Ti_2B_2O_2$ at the Fermi level is higher compared to ml-$Ti_2B_2Cl_2$, which suggests that oxygen terminations enhance the electrical conductivity of the material.

The theoretical results presented herein support our experimental findings. While the evaluation of redox potential of the elements in their elemental state is shown to be insufficient for predicting MS etching of $Ti_2InB_2$, the recently proposed framework developed for MAX phases [23] accurately assesses the reactivity of MAB phases in MS. Moreover, we anticipate that the approach can be extended to other MAB phases and layered compounds beyond borides.

To further evaluate our theoretical model, we investigated the reaction between $Ti_2InB_2$ and $CuCl_2$. The high redox potential of $Cu^{2+}/Cu$ makes this salt an effective Lewis acid to etch different A-elements in MAX phases [25]. Different attempts of using $CuCl_2$ salt to etch MAB phases, were unsuccessful in achieving MBenes [16–19]. According to our model, shown in Figure 3a, the calculated reaction free energies for all four possible scenarios at 900K are comparable at low salt-to-MAB molar ratios, while complete disintegration becomes the most favorable outcome at higher ratios. Moreover, the predicted reaction energies are significantly lower (more negative) than those for the $ZnCl_2$ case, which can be correlated with the higher reactivity of the $CuCl_2$ salt.

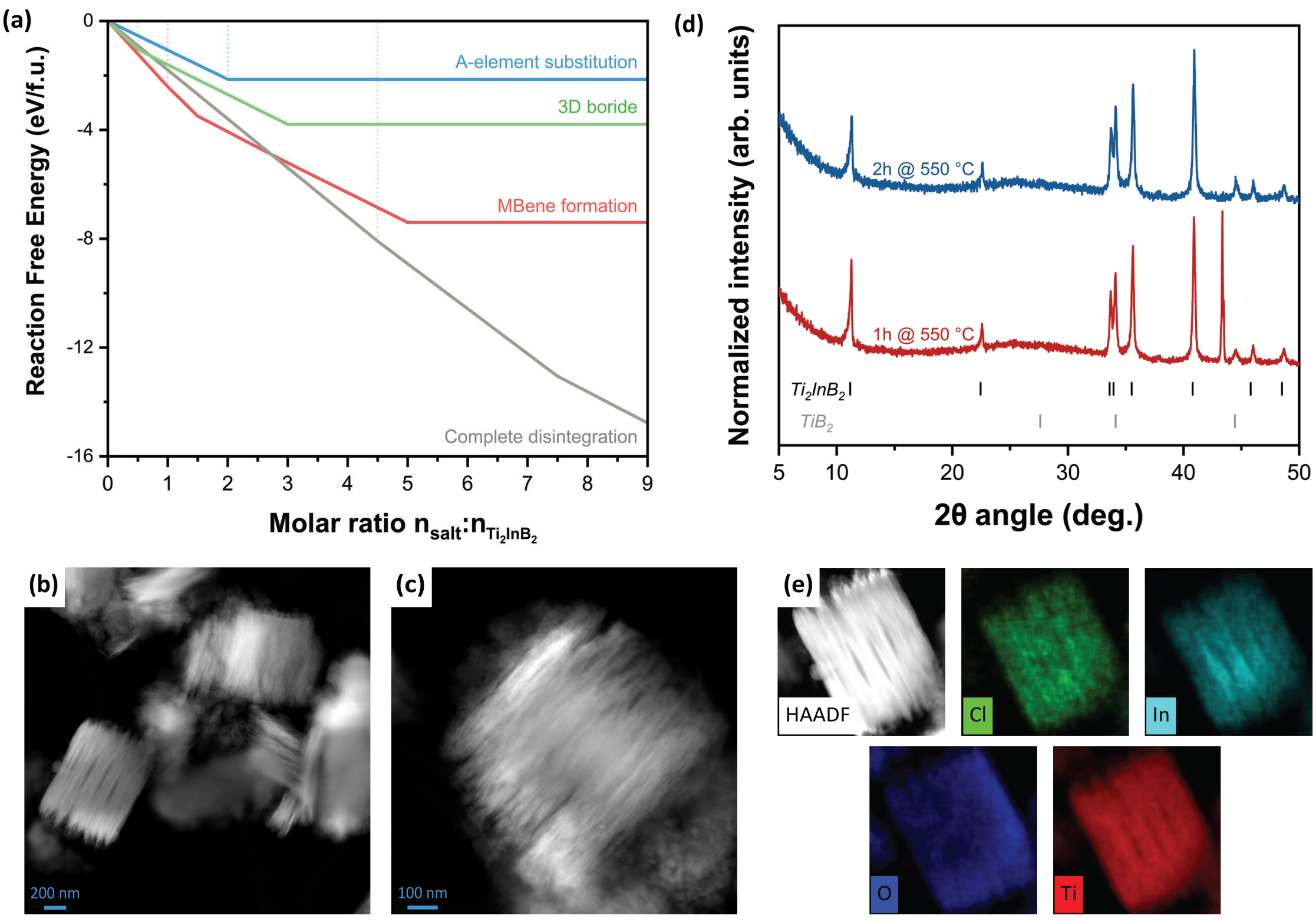

Figure 3: (a) Gibbs free energies as a function of the salt-to-MAB molar ratio for $CuCl_2$ as etchant. (b) XRD of samples obtained after MS etching, (c, d) high-magnification HAADF-STEM images, and (e) EDX maps of a representative particle.

Initial experiments were carried out at 550 °C for different reaction times. The XRD patterns obtained after MS reactions are similar to the 3D precursor (Figure 3b), but multilayer particles can be seen in the HAADF-STEM images (Figure 3c-d). STEM-EDX analysis confirms the coexistence of In, Cl, and O within the particles, with approximately two-thirds of In being removed (In/Ti ≈ 0.15). However, different from etching attempts in $ZnCl_2$, the chlorine content remains negligible (Cl/Ti ≈ 0), and oxygen is distributed throughout the entire particle rather than being confined to the particle edges (sheet edges) (Figure 3e). It should be noted that EDX does not resolve the chemical state of the element, and the detected oxygen may originate from surface terminations on $Ti_2B_2T_x$ and/or from partial oxidation of Ti-containing species. Furthermore, the mass loss after etching was approximately 10%, considerably lower than the ~50% observed for the $ZnCl_2$ process. These observations may suggest structural decomposition of the MAB phase, in which In and B species (or their reaction products, e.g., In–Cl compounds) may be partially removed during processing, while Ti-containing species are oxidized. This is consistent with the thermodynamic predictions, but the similar reaction energies across all processes highlight the complexity of the $CuCl_2$-based reaction and indicate multiple competing processes.

MS-derived MXenes are sometimes formed through an intermediate step, in which the original A-layer is substituted by a metal species supplied by the molten salt. [22,24] Recently, Zn-intercalation was also indicated when $Ti_2InB_2$ was reacted with $ZnCl_2$ within a short time frame [10]. This is different from both experimental and theoretical results shown herein, where no A-layer substitution in the MAB phase is seen prior to the ml-MBene formation. As can be seen in Figure S13, after experiments performed at 0.5h at 600°C, there is no indication of Zn-intercalation, regardless of the salt ratio employed.

To investigate this question in detail, we monitored the pathway of the $Ti_2InB_2$ reaction in molten $ZnCl_2$, by *in situ* synchrotron XRD. Figure 4a displays the XRD patterns recorded every 40 s with an acquisition time of 10 s, along the temperature profile encompassing a 10 °C·min$^{-1}$ heating ramp to 600 °C, followed by a 4 h plateau, and then further increase to 700 °C for 75 min before cooling down.

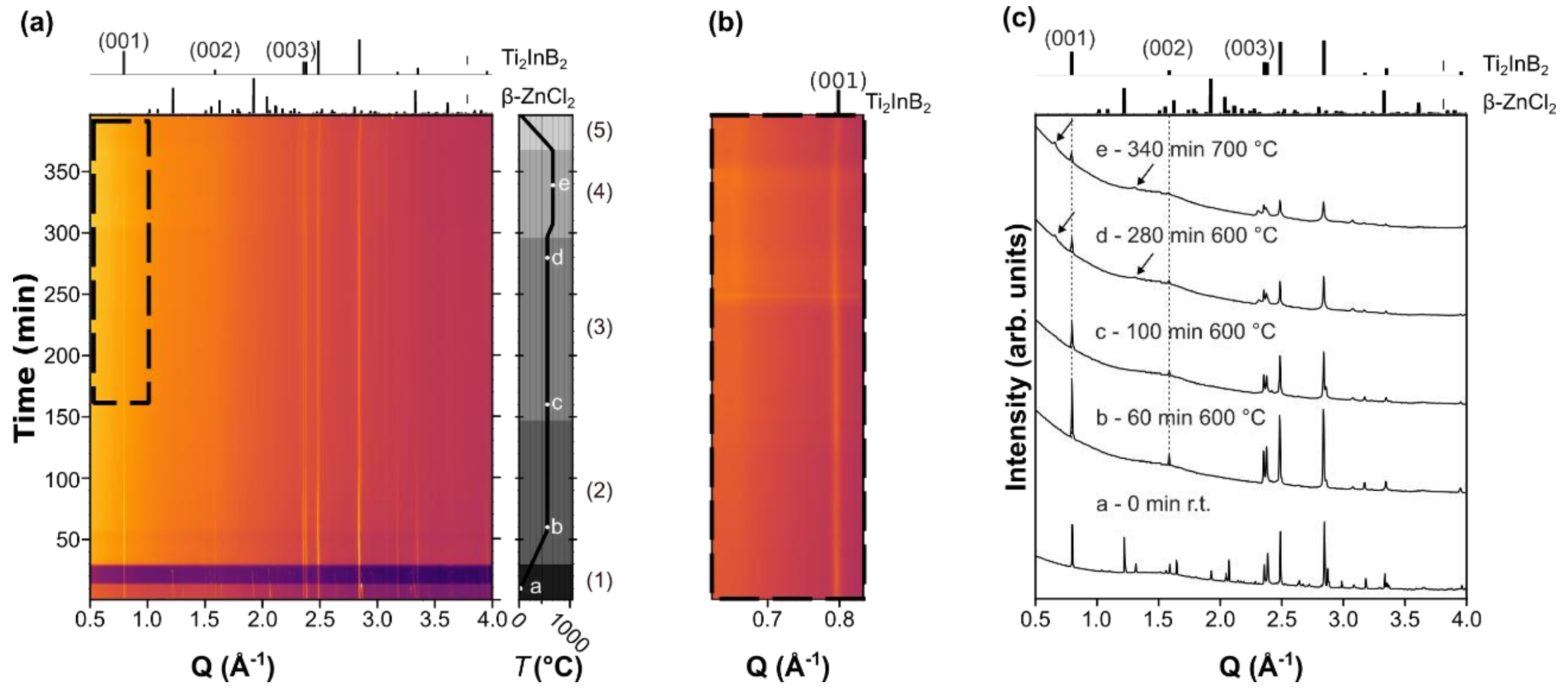

Figure 4: *In situ* monitoring of $Ti_2InB_2$ reactivity in molten $ZnCl_2$ using 1 $Ti_2InB_2$:3 $ZnCl_2$ molar ratio. (a) *In situ* XRD patterns recorded during the reaction and displayed as a heatmap along the corresponding temperature profile; (b) corresponding zoom in stages (3) to (5) for low Q values. (c) Selected XRD patterns corresponding to the main areas highlighted in the temperature profile in (a). Room-temperature reference patterns are also displayed at the top of (a) and (c). Sudden changes of overall intensity in (a) between ~20 and ~30 min are ascribed to a change in the thickness of the sample, while salt grains coalesce when approaching the melting point.

Five stages can be distinguished (Figure 4a, enlarged in Figure S14), with characteristic XRD patterns displayed in Figure 4c (enlarged in Figure S15). Stage (1) corresponds to the initial mixture made of $Ti_2InB_2$ and β-$ZnCl_2$, a partially hydrated zinc chloride allotrope (Figure S15). [34] The second stage starts when zinc chloride melts. The melting point of zinc chloride is observed after 30 min at 325 °C when its XRD peaks disappear (Figure S16), in agreement with our previous work [22]. Above this temperature, a scattering signal is observed below 1.8 $Å^{-1}$ where the background intensity is increased due to the liquid medium (Figure 4a and Figure S16). An additional broad signal between 1.2 and 1. 8 $Å^{-1}$ is ascribed to the capillary walls. $Ti_2InB_2$ is then the only crystalline phase detected. It is dispersed in the molten salt. Within the second stage (2), from the beginning of the plateau at 600 °C and for 100 min, the {001}, {002}, and {003} reflections of $Ti_2InB_2$ at 0.796, 1.585, and 2.377 $Å^{-1}$ (interlayer distance 7.913 Å) are not significantly shifted. Thereby, during this step, the distance between the titanium boride layers is retained. At 280 min, after 2 h of plateau, the third stage starts, with the sudden appearance of two peaks at 0.661 and 2.311 $Å^{-1}$ and the concomitant decrease of the intensity of the (001) $Ti_2InB_2$ reflection. The 0.661 Å-

[1] (9.501 Å) closely agrees with the interlayer distance measured *ex situ* and calculated for the chloride terminated ml-MBene (Figure S17).

The absence of a progressive shift of this reflection indicates a biphasic transformation from the MAB phase to the ml-MBene, as already observed in the transformation of $Ti_3ZnC_2$ to $Ti_3C_2Cl_2$ [22,24]. We note, however, that we do not observe any significant structural change before the formation of the ml-MBene, in agreement with *ex situ* results. This transformation for a MAX phase like $Ti_3AlC_2$ occurs in a two-step process, with the intermediate formation of a Zn-exchanged intermediate (e.g. $Ti_3ZnC_2$). A similar replacement of In by smaller Zn atoms would result in a decrease of the interlayer distance, hence an upward shift of the corresponding diffraction peaks, which is not observed. These *in situ* data then show that, at the temporal resolution of the analysis, there are no intermediate compound formed. This behavior contrasts with that observed in some cases for MXenes synthesized by MS-etching of MAX phases.

Likewise, no transient expansion of the interlayer space was identified during *in situ* measurements, which rules out a reaction mechanism through swelling, in agreement with previous data on $Ti_3AlC_2$ reacted in molten $ZnCl_2$ [22,35] and later work on the same compound [36]. We also did not observe any significant decrease in the intensities of the $Ti_2InB_2$ diffraction peaks, nor the broadening of these peaks before the appearance of the ml-MBene reflection (Figure S17). In agreement with our previous work on $Ti_3AlC_2$ [22], this suggests that the reaction does not proceed through an amorphous intermediate mentioned in other reports [36]. Two possible scenarios can then be considered: (1) swelling occurs to enable redox reaction between $ZnCl_2$ and In atoms up to the core of the particles, but it is too fast to be captured with the time resolution of our *in situ* set-up (10 s acquisition time); (2) as the reaction proceeds at the edge of the layers on the surface of the particles, indium vacancies form at the edge, which triggers In atom diffusion from the core to the edge of the layer and inward diffusion of chlorine atoms. These two pathways cannot be distinguished in the present work and require a more targeted investigation.

To favor the ml-MBene, we have further increased the temperature to 700 °C in a fourth stage, but no significant change of the relative intensities could be detected. In the fifth stage, the reaction medium is cooled down, which is accompanied by an upward shift of the reflections due to thermal dilatation and the appearance of new peaks caused by the recrystallization of the melt (Figure S18). These peaks can be tentatively assigned to $ZnCl_2$ polymorphs, Zn, and $InCl_3$. The latter is consistent with a galvanic etching reaction in which $ZnCl_2$ oxidizes In atoms to chloride species. Metallic zinc might also be present, but the corresponding peaks could overlap with those of $Ti_2InB_2$, and the ml-MBene.

Finally, the ml-$Ti_2B_2Cl_x$ powder was evaluated for Li-ion battery applications. Its electrochemical characterization (Figure S19) revealed a capacitor-like behavior with no distinct redox peaks in the cyclic voltammetry curves. The material delivered an initial discharge capacity of ~226 $mAhg^{-1}$, and after a slight decrease in capacity within the first hundreds of cycles, it started to increase gradually. The coulombic efficiency on the first cycle was 42%, likely as a result of the formation of the solid electrolyte interphase (SEI) layer, and then increased to nearly 100%, remaining stable thereafter. Long-term cycling tests at 0.5 and 1 $Ag^{-1}$ current densities demonstrated a stable performance with discharge capacities of 157 and 135 $mAhg^{-1}$, respectively. This performance is similar to $Ti_2InB_2$ [37], better than $Nb_2SnC$ [38] and $Ti_2SC$ [39] MAX phases, and slightly below MS-$TiBT_x$ [10] (see Table S5). This performance may be further improved by optimization and by tuning of surface terminations, such as increasing the concentration of oxygen terminations, previously shown to be more promising in both MXenes and MBenes materials for energy storage applications [10,40,41].

## CONCLUSIONS

The synthesis of multilayer MBene, $Ti_2B_2Cl_x$ ($1.1<x<1.4$) was achieved via molten salt etching of $Ti_2InB_2$ in $ZnCl_2$. The 3D $Ti_2InB_2$ precursor was prepared in a 6h annealing plateau solid-state reaction, significantly faster than previously reported (6 days) [20]. After its reaction with $ZnCl_2$ salt, EDX and EELS confirmed complete removal of In accompanied by the insertion of Cl terminations with coverages $1.1<x<1.4$, while STEM identified an expanded interlayer spacing, corroborated by an increased (002) d-spacing as shown by XRD. Further, *in situ* XRD indicated a direct biphasic transformation from $Ti_2InB_2$ to ml-MBene, distinct from MXene reactions that usually present an intermediate compound after A-layer exchange [22]. DFT calculations identified the thermodynamic driving force behind this process, with $Ti_2B_2Cl_2$ exhibiting the lowest (and only negative) reaction free energy among possible pathways and predicting the A-element substitution resulting in $Ti_2ZnB_2$ to be thermodynamically unfavorable. Initial electrochemical characterization for Li-ion batteries showed discharge capacities similar to or superior to MAX phases and other borides. These experimental and theoretical results advance the prediction and synthesis of MBenes prediction and synthesis, while providing a framework for targeting unexplored MBenes accessible via MS reaction pathways.

## METHODS

### Computational Details

The DFT [42,43] calculations were performed using Vienna *Ab initio* Simulation Package (VASP) [44,45] within a plane-waves basis set expanded to a kinetic energy of 520 eV. The rev-vdW-DF2 functional [30] was used to model the exchange-correlation interactions, while the Projector Augmented Wave (PAW) method [46,47] was used to describe the electron-ion interaction. The structure relaxations used Monkhorst–Pack grids [48] to sample the Brillouin zone, with a k-point density of 1/6 $Å^{-1}$. Energy and force criterion for self-consistency were set as $10^{-7}$ eV/atom and $10^{-3}$ $eVÅ^{-1}$. Initially, both non-magnetic and ferromagnetic spin configurations were considered but they converged to the non-magnetic configuration. Therefore, all results shown here do not include any magnetic interactions.

Phonon dispersion calculations were computed using Phonopy code [49,50] on supercells of at least 10 Å in each direction. Instabilities around the Γ point for 2D materials were recently associated with the lack of rotational invariance in the harmonic lattice Hamiltonian and non-vanishing external stress from DFT calculations [51,52]. These may be a result of incomplete basis sets, insufficient Brillouin zone sampling or numerical stability of algorithms, especially in the direction of vacuum and/or weak van der Waals forces between layers and interactions from the periodic images. To ensure rotational and translational invariances, the phonon spectrum was calculated by applying constraints to the force constants. The rotational and translational sum rules according to the Born-Huang condition and the Huang invariances [53], were enforced using the hiPhive Python package [54].

The molten salt and MAB phase reaction can be written as:

$$MAB + \frac{n_{salt}}{n_{MAB}} A'Ha_y \rightarrow product \tag{1}$$

where $n_{salt}$ and $n_{MAB}$ are the molar amounts of salt and MAB, respectively, $A'Ha_y$ is the salt (here $ZnCl_2$), consisting of a halogen ($Ha_y = Cl_2$) and a metal ($A' = Zn$).

To assess the thermodynamics of this reaction, we used a recently developed methodology applied to MAX phases and MXenes [23], which maps different competing reactions by constraining the product term in Eq. {1} to a set of possible compounds that are specific to each etching scenario. Four distinct scenarios were compared: (i) A-element substitution, where the A-element of the MAB phase is replaced by $A'$ from the salt; (ii) MBene formation, in which the A-element is etched, and halogens from the salt bind to the surface of the resulting MBene, (iii) complete disintegration, where the MAB phase fully decomposes into salts or its constituent elements; (iv) 3D boride formation, where the A-element is etched, and the MBene layers collapse into a 3D boride. These scenarios are schematically summarized Figure 3a.

For each scenario and salt-to-MAB ratio, the total free energy of the products was minimized by using linear combinations of the formation energies of a constrained set of products specific to each scenario and maintaining the overall reactant composition (referred to as process-specific phase diagrams). The reaction free energy is then calculated as the difference between the free energy of the products and that of the reactants.

Regardless of the scenario evaluated, there are common phases that are always allowed at the product term: the original MAB phase, elemental $A'$, salts between $A$ and $Ha$, and salts between $A'$ and $Ha$ with lower oxidation state of the metal compared to the initial etching salt ($A'Ha_y$). In addition, the following process-specific products are considered in the different scenarios:

(i) A-element substitution: The substituted MAB phase, $MA'B$;

(ii) MBene formation: The resulting MBene, $M_nB_nHa_2$;

(iii) Complete disintegration: Salts between $M - Ha$, and salts between $B - Ha$ , as well as elemental forms of $M$, $A$ and $B$;

(iv) 3D carbide/nitride formation: The $MB$ boride.

For each compound considered in reaction {1}, the Gibbs free energy of formation was calculated as:

$$\Delta G(material; T) = G(material; T) - \sum G(element) \qquad \{2\}$$

where $G(material; T)$ is the Gibbs free energy of the material at a certain temperature $T$ and $\sum G(element)$ is the sum of the free energies of its elemental constituents at standard conditions (298.15 K and 1 bar). Since both calculated and tabulated formation free energies were employed [55,56], a consistent treatment of reference states is needed to ensure comparability between calculated and tabulated energies.

The Gibbs free energies of formation for MAB phases, substituted MA'B phases, MBenes and 3D metal borides, were derived from first-principles calculations. The free energies of the solid-state materials were computed within the harmonic approximation, combining vibrational free energies (entropy and enthalpy from phonon calculations) with electronic enthalpy (DFT total energy). To simulate removal of the A-layer, resulting in a collapsed 3D structure, we chose to include TiB (*Pnma*), since it is experimentally known and has a lower DFT-calculated energy (by 92 meV/atom) compared to the collapsed 3D structure generated using the procedure described in ref [23]. For the gas phase, $Cl_2$, contributions from translational, rotational and vibrational enthalpy were included alongside tabulated total entropy values [57]. For all other materials, such as the salts and other products from etching, Eq {2} employed tabulated free energies of both the material in question and its elemental constituents. The use of tabulated values ensures the use of the correct salt phase. For example, the formation free energy of $ZnCl_2$ would correspond to the solid phase if the MS reaction were evaluated at 400 K, while the liquid phase would be considered at 900 K.

## Experimental Details

***Synthesis of $Ti_2InB_2$:*** Elemental powders (see SI, Table S6) were mixed using a mortar and pestle and annealed at 1100 ˚C for 6 hours in a tube furnace under 5 sccm Argon flow (heating and cooling rates of 3 ˚C/min). Following ref [20], 50% excess amounts of both Ti and In were added to achieve higher sample purity. After the solid-state reaction, the powder was immersed in diluted hydrochloric acid (2 mol $L^{-1}$) at room temperature for 24 hours to remove Ti–In impurity phases (e.g., $Ti_3In$, $Ti_3In_4$, $Ti_{2.24}In_{1.76}$). The only phases remaining after this cleaning procedure were $Ti_2InB_2$ and $TiB_2$.

***Synthesis of ml-$Ti_2B_2Cl_2$:*** The synthesized $Ti_2InB_2$ powder was mixed with $ZnCl_2$ salt, cold-pressed using a load corresponding to a stress of 50 MPa, and the resulting pellets were sealed inside a quartz tube under vacuum (pressure lower than $5\times10^{-5}$ mbar). Different synthesis conditions (Fig. S5) were tested, with optimum conversion achieved using 1 $Ti_2InB_2$:4 $ZnCl_2$ molar ratio and annealing at 600 ˚C for 5 hours (with heating and cooling rates of 5 ˚C $min^{-1}$). After the reaction, the pellet was first cleaned with deionized water to remove unreacted salts and then with diluted HCl (1 mol $L^{-1}$) to potentially remove the excess of Zn metal. Finally, the washed sample was filtered and crushed to achieve a powder.

## Materials characterization

X-ray diffraction (PANalytical X'Pert Pro diffractometer, Cu $K_\alpha$ radiation source, X'Celerator line detector) was used to probe the crystal structure, with divergence and anti-scatter slits of 1/2° each on the incident side and anti-scatter slit of 5 mm on the receiving side. Structural parameters and phase content were obtained through a Rietveld refinement procedure [58,59] using the FULLPROF software [60]. Thompson-Cox-Hastings pseudo-Voigt with an axial divergence asymmetry function was used to model peak shapes, and instrument parameters were determined using a NIST $LaB_6$ standard material. For Rietveld analysis, the step size and time per step used to record the XRD patterns were 0.0084° and 32 s, respectively. The refined parameters were scale factors, zero offset, five background parameters, lattice parameters, Y profile, asymmetry parameters, overall thermal factor, preferred orientation and atomic positions.

Materials morphology and chemical composition were evaluated using Scanning Electron Microscopy, SEM (Zeiss Sigma 300), combined with an energy dispersive X-ray spectrometer (Oxford Instruments X-Max 80 $mm^2$ SDD detector). For each material, at least 20 particles were evaluated to acquire statistics.

To investigate the atomic structure of the ml-MBene material/derivative, high-angle annular dark-field (HAADF) scanning transmission electron microscopy (HAADF-STEM) imaging, EDX and electron energy loss spectroscopy (EELS) analyses were performed using the Linköping double-corrected and monochromated FEI Titan$^3$ (S)TEM electron microscope, operated at 300 kV. For analysis, the particles were dispersed onto standard lacey amorphous carbon support films suspended by Cu grids. High-resolution HAADF-STEM images were recorded using an optimized 30 mrad probe convergence angle, which provided sub-Å size probes and ~80 pA probe current. EDX analysis were performed employing embedded SuperX EDX detector. EELS spectra were collected using Gatan GIF Quantum ERS post column imaging filter in dual EELS mode and quantified via built-in functions in Gatan Digital Micrograph.

*In situ* XRD experiment was carried out on the DiffAbs beamline at the French SOLEIL Synchrotron facility. The experiment was performed in transmission mode at an energy of 18.0 keV (wavelength = 0.689 Å). The energy was tuned thanks a double crystal monochromator composed of two Si (1,1,1) crystals. The spot size had a diameter of around 300 microns at sample position. The CIRPAD detector used for the experiment was a circular 2D detector covering a 140° detection angle in a single acquisition [61]. It is composed of 20 hybrid pixel modules (XPAD) separated by inter-module spacing which do not detect X-rays. The diffractogram was the result of the average of 2 images (by a tilt of 3° of the detector) to fill in this inter-module gaps.[22,61] The acquisition time for each separate image is 3 s, which yields an acquisition time of 10 s for each reconstructed pattern. A custom-made capillary oven controlled the heating of the reaction mixture, placed in a 1 mm diameter fused silica capillary, while recording the X-ray diffraction. The capillary was filled in an argon-filled glovebox, flame-sealed under vacuum outside of the box without contact with air, enabled by a specially designed sample holder.

For evaluation of the electrochemical properties, ml-$Ti_2B_2Cl_x$ (active material, 70 wt%) was mixed with conductive agent Super P (20 wt%) and PVDF (10 wt% each). This mixture was dispersed in N-methyl-2-pyrrolidone (NMP) and homogenized using a planetary mixer (2000 rpm, 1 h) to form a uniform slurry. The slurry was then applied onto a copper foil current collector (thickness: 8 μm) using a doctor blade coater with a 150 μm gap. The coated electrodes were vacuum dried at 60°C for 12 h and then cut into discs with a diameter of 12 mm. The active material mass loading was controlled within the range of 1.12–2 mg $cm^{-2}$. CR2032 coin cells were assembled in an argon-filled glove box ($H_2O$ < 1 ppm and $O_2$ < 0.1 ppm). The working electrode consisted of the $Ti_2B_2Cl_2$ anode material, the counter electrode was a lithium metal foil (diameter: 15.6 mm, thickness: 0.6 mm, Kelude), and the separator was PP/PC/PP film (25 μm thickness). The electrolyte used was 1 M $LiPF_6$ in a mixture of ethylene carbonate (EC), ethyl methyl carbonate (EMC), and dimethyl carbonate (DMC) (volume ratio 1:1:1) with 10 wt% fluoroethylene carbonate (FEC) additive. Galvanostatic charge/discharge tests were performed using a Shenzhen Neware CT-4008-5V6A test system within a voltage window of 0.01–3.0 V (current density gradient: 0.01–5 $Ag^{-1}$). The test termination criterion was when the discharge capacity decayed to 80% of the initial cycle or the Coulombic efficiency fell below 98%. Cyclic voltammetry (CV) was conducted using a Wuhan Corrtest CS Studio6 electrochemical workstation within a voltage range of 0.01–3.0 V (vs. $Li/Li^+$) at a scan rate of 0.1 $mVs^{-1}$ (initial 3 cycles for activation).

## ACKNOWLEDGEMENTS

We acknowledge support from the Knut and Alice Wallenberg (KAW) Foundation for Scholar (2023.0250) and Project funding (KAW 2020.0033), from the European Union (ERC, MULTI2D, 101087713 and ERC GENESIS, 864850), and from the Swedish Government Strategic Research Area in Materials Science on Advanced Functional Materials at Linköping University (Faculty Grant SFO-Mat-LiU No. 2009-00971). This work received funding from the French Investments for the Future Program (PIA) as part of the France 2030 Initiative operated by the National Agency for Research (ANR), under the project MADNESS (grant agreement 23-PEXD-0011). The KAW Foundation is also acknowledged for the support of the electron microscopy laboratory at Linköping University, and the Swedish Research Council (VR) for access to ARTEMI, the Swedish National Infrastructure in Advanced Electron Microscopy (2021-00171 and RIF21-0026). The experiments at DiffAbs beamline of synchrotron SOLEIL have been supported by SOLEIL as the user proposal 20240574. The computations were enabled by resources provided by the National Academic Infrastructure for Supercomputing in Sweden (NAISS) at the National Supercomputer Centre (NSC), partially funded by the Swedish Research Council through grant agreement no. 2022-06725.